# Spin-orbit interactions of transverse sound

Shubo Wang[1,*], Guanqing Zhang[2], Xulong Wang[2], Qing Tong[1], Jensen Li[3], and Guancong Ma[2,*]

[1] Department of Physics, City University of Hong Kong, Tat Chee Avenue, Kowloon, Hong Kong, China

[2] Department of Physics, Hong Kong Baptist University, Kowloon Tong, Hong Kong, China

[3] Department of Physics, The Hong Kong University of Science and Technology, Clear Water Bay, Hong Kong, China

*Correspondence should be addressed to:
Shubo Wang (shubwang@cityu.edu.hk)
Guancong Ma (phgcma@hkbu.edu.hk)

## Abstract

Spin-orbit interactions (SOIs) endow light with intriguing properties and applications such as photonic spin-Hall effects and spin-dependent vortex generations. However, it is counterintuitive that SOIs can exist for sound, which is a longitudinal wave that carries no intrinsic spin. Here, we theoretically and experimentally demonstrate that airborne sound can possess artificial transversality in an acoustic micropolar metamaterial and thus carry both spin and orbital angular momentum. This enables the realization of acoustic SOIs with rich phenomena beyond those in conventional acoustic systems. We demonstrate that acoustic activity of the metamaterial can induce coupling between the spin and linear crystal momentum **k**, which leads to negative refraction of the transverse sound. In addition, we show that the scattering of the transverse sound by a dipole particle can generate spin-dependent acoustic vortices via the geometric phase effect. The acoustic SOIs can provide new perspectives and functionalities for sound manipulations beyond the conventional scalar degree of freedom and may open an avenue to the development of spin-orbit acoustics.

Spin and orbital angular momentum (OAM) are intrinsic properties of classical waves. Spin is associated with circular polarization (vector degrees of freedom) of waves and is characterized by the local rotation of a vector field. OAM originates from the spatial phase gradient (scalar degree of freedom) of waves and manifests as a helical wave front[1]. The couplings between spin and OAM, referred to as spin-orbit interactions (SOIs), can give rise to intriguing phenomena and applications in optics[2–8], such as photonic spin-Hall effect[9,10] and spin-dependent vortex generation[11,12]. SOIs are unique to transverse waves such as light and are absent for longitudinal waves. This is because although longitudinal waves such as airborne sound can carry OAM [13–17], they are spin-0 in nature. Recent studies show that an engineered sound field can possess a locally rotational velocity field **v** that may be regarded as acoustic spin[18–20], similar to electric spin deriving from the local rotation of electric field. Such an acoustic spin can emerge locally in nonuniform acoustic fields[19,20] and has recently been observed in experiments[18,21]. In a homogenous medium, however, the spatial integration of acoustic spin density for a localized wave must vanish, in agreement with its spin-0 nature[19]. Despite this new discovery of acoustic spin, SOIs remain beyond reach in sound, a fact that mainly owes to the lack of degrees of freedom. In other words, sound is characterized by a scalar pressure field $p$ and a vector velocity field **v**, whereas light is characterized by two vector fields **E** and **H**.

In this work, we show that airborne sound can behave as a transverse wave with well-defined polarization in an acoustic metamaterial that goes beyond the Cauchy elasticity and follows a micropolar elasticity theory[22]. Unlike previous spin-sustaining acoustic fields[18,19,21], the transverse sound is spin-1 in nature and carries the properties of elastic waves. It is characterized by two types of vector-field degrees of freedom, i.e., a velocity field and a microrotation field. The acoustic activity of the metamaterial can induce coupling between the velocity and microrotation fields, which can be considered an analogue of chirality in electromagnetism (i. e., optical activity). Such a material property has recently been realized in elastic wave systems[23–26] but is so far missing in acoustic wave systems. We theoretically and experimentally demonstrate two types of acoustic SOIs in momentum space and in real space, respectively. In the momentum space, the acoustic activity induces the coupling between spin and linear crystal momentum **k** and enables the chirality-induced negative refraction, which was previously possible only in optical metamaterials[27,28]. In the real space, scattering of the circularly polarized transverse sound by a dipole particle can generate a sound vortex with a topological charge determined by the acoustic spin.

**Results**

**Transverse sound**

The longitudinal nature of airborne sound ($\boldsymbol{\nabla} \times \mathbf{v} = 0$) dictates that the velocity field **v** aligns with the direction of wave vector **k** in general. However, this is not necessarily true when sound is confined in a closed space. Consider a one-dimensional (1D) lattice stacked along *z* axis with a unit cell shown in Fig. 1(a). The unit cell consists of a cylindrical resonator with eight internal blades segmenting the air to achieve subwavelength resonance, as indicated by the blue arrows. The resonators are sequentially connected by four tubes. All solid-air interfaces are regarded as sound-hard boundaries. The resonator supports two degenerate and orthogonal dipole resonances with pressure eigenfields shown in Fig. 1(b). The positive and negative pressure (indicated by the red and blue colors, respectively) induces an in-plane velocity field that is perpendicular to the propagating direction of sound (i.e. *z* axis). This corresponds to the oscillating dipole moments $\mathbf{p}_x$ and $\mathbf{p}_y$, where the positive (negative) charge corresponds to the positive (negative) pressure and the yellow arrow denotes the velocity field. Next, we break the spatial inversion symmetry by twisting the resonator geometry with respect to *z* axis, as shown in Fig. 1(c). The degeneracy of $\mathbf{p}_x$ and $\mathbf{p}_y$ is removed, and the resonator supports two chiral eigenmodes $\mathbf{p}_x + i\mathbf{p}_y$ and $\mathbf{p}_x - i\mathbf{p}_y$, corresponding to a left-handed circularly polarized (LCP) dipole and a right-handed circularly polarized (RCP) dipole, respectively, as shown in Fig. 1(d). Thus, the collective excitations of the acoustic dipoles in Fig. 1(b) and 1(d) will give rise to linearly polarized and circularly polarized transverse sounds propagating in *z* direction, respectively.

To verify this, we use 3D printing to fabricate both the 1D achiral and chiral lattices, each with 24 unit cells, as shown in Fig. 2(a). In Fig. 2(b), we show the cutaway views of the two types of unit cell, where the internal blades are colored to clearly show their orientations. The experimentally measured band structures of the achiral and chiral lattices are shown in Fig. 2(c) and 2(d), respectively. The solid red lines denote full-wave numerical results calculated using a finite-element package COMSOL (see Methods). Excellent agreement between the experimental and numerical results is seen. The 1st band that extends to the static limit corresponds to a monopole mode, which has almost identical characteristics for both the chiral and achiral lattices. The 2nd and 3rd bands are the aforementioned transverse dipole modes, which are degenerate for the achiral lattice (Fig. 2(c)) but split into two bands for the chiral lattice (Fig. 2(d)) due to inversion symmetry breaking. The modes of the 2nd and 3rd bands are LCP and RCP, respectively. To obtain intuitive pictures of the transverse modes, we calculated the averaged velocity (near $k_z = 0$) in each unit cell and plot it in Fig. 2(e) and 2(f). Figure 2(e) shows the velocity field for the achiral lattice with 25 units. As seen, the sound is linearly polarized along *y* direction with a wavelength much larger than the unit-cell dimension. Figure 2(f) shows the velocity field for the 2nd band of the chiral lattice, which clearly represents an LCP transverse sound. These confirm the transverse nature of the sound in the 1D lattices.

**Micropolar metamaterial with acoustic activity**

The above physics can be extended to the three-dimensional (3D) metamaterial with the unit cell shown in Fig. 3(a). The unit cell consists of three chiral resonators mutually connected with tubes, as shown in Fig. 3(b). The numerically calculated band structure of the metamaterial is shown in Fig. 3(c). The three bands enclosed by the red rectangle derive from the dipole modes. The upper and lower bands correspond to the RCP and LCP transverse modes, respectively, and the middle band corresponds to a longitudinal mode. The inset at the left corner of Fig. 3(c) shows the pressure eigenfield of the LCP mode at a time. In Fig. 3(d) and 3(e), we plot the isofrequency contours of the LCP band in $k_x$-$k_y$ and $k_z$-$k_M$ planes, respectively. The contours are approximately circles for $k < 0.15\pi/a$, which indicates that the mode is isotropic near the $\Gamma$ point. The isotropic dispersions of the transverse modes are protected by time-reversal symmetry and chiral cubic symmetry[29]. At the frequencies of the transverse modes, the unit cell is subwavelength ($\sim 0.23\lambda$). Thus, the metamaterial is macroscopically isotropic and homogeneous, and its material properties can be described by an effective medium theory. Remarkably, the emergence of wave transversality in the metamaterial implies the existence of a non-zero shear modulus for the effective medium, which is counterintuitive since air does not generate shear forces. Here, the striking properties of non-vanishing shear modulus are induced by the transverse motion of sound enforced by the resonators with twisted internal blades. The existence of a non-zero shear modulus indicates that the metamaterial is equivalent to an elastic medium, and the airborne sound behaves like an elastic wave with well-defined spin[30]. Because of its microscopic twisting feature, the metamaterial cannot be described by conventional effective medium theory based on Cauchy elasticity, which assumes symmetric stress and strain. Instead, micropolar elasticity (i. e., Cosserat elasticity)[22], which is a high-order extension of Cauchy elasticity, can be employed to accurately characterize its unusual properties.

The micropolar elasticity assigns three rotational degrees of freedom to each material point in addition to the three linear degrees of freedom associated with displacement[22,31–33]. Each point is thus characterized by a displacement vector field **u** and a microrotational vector field **ϕ**. Using Einstein summation convention, the deformation of the medium can be expressed as: $\varepsilon_{ij} = \partial u_j/\partial x_i - \epsilon_{ijk}\phi_k$; $\kappa_{ij} = \partial \phi_j/\partial x_i$, where $\varepsilon_{ij}$ is the asymmetric strain tensor, $\kappa_{ij}$ is the curvature tensor characterizing the relative microrotation between neighboring points, $\epsilon_{ijk}$ is the Levi-Civita symbol, and *i, j, k* iterate the Cartesian coordinates. For our micropolar metamaterial, the corresponding effective medium is characterized by the constitutive relations: $\sigma_{ij} = C_{ijkl}\varepsilon_{kl} + B_{ijkl}\kappa_{kl}$, $m_{ij} = B_{klij}\varepsilon_{kl} + D_{ijkl}\kappa_{kl}$, where $\sigma_{ij}$ and $m_{ij}$ are the asymmetric force stress tensor and couple stress tensor, respectively [25,34]. $B_{ijkl}, C_{ijkl}$ and $D_{ijkl}$ are the elastic constitutive tensors of the form $X_{ijkl} = X_1\delta_{ij}\delta_{kl} + X_2\delta_{ik}\delta_{jl} + X_3\delta_{il}\delta_{jk}$ with $X = B, C, D$ and $\delta_{ij}$ being the Kronecker delta. Notably, $B_{ijkl}$ is a pseudo-tensor that characterizes the chirality of the medium and it changes sign under spatial inversion. Thus, the micropolar metamaterial possesses chirality that corresponds to the acoustic counterpart of optical activity[35]. Such a property has recently been realized in elastic metamaterials[23–25] but has no acoustic counterpart to date. It is different from the Willis-type bianisotropy, in which the stress-strain couples with momentum-velocity[36–41].

The propagation of the transverse sound is governed by the conservation of linear and angular momenta: $\partial \sigma_{ji}/\partial x_j = \rho \partial^2 u_i/\partial t^2$; $\partial m_{ji}/\partial x_j + \epsilon_{ijk}\sigma_{jk} = j\partial^2\phi_i/\partial t^2$, where $\rho$ is the mass density and $j$ is the microinertia density (i.e. micro momentum of inertia per unit volume).

Assuming the time-harmonic displacement eigenfield $u_i = U_i e^{i k_i x_i - i\omega t}$ and microrotation eigenfield $\phi_i = \Phi_i e^{i k_i x_i - i\omega t}$, the dispersion relations of the dipole modes near the $\Gamma$ point (retained the lowest order of $k$) can be obtained as (see Methods): $\omega_T^{\pm} = \omega_0 \pm vk$ and $\omega_L = \omega_0 + \tau k^2$, where $\omega_0 = \sqrt{2(C_2 - C_3)/j}$, $k = |\mathbf{k}| = \mathbf{k}/\hat{\mathbf{k}}$, $v = (B_2 - B_3)/\sqrt{2j(C_2 - C_3)}$, $\tau = (D_1 + D_2 + D_3)/\sqrt{8j(C_2 - C_3)}$, and the subscripts "T" and "L" denote the transverse and longitudinal modes, respectively. It is seen that microrotation significantly impacts both the transverse and longitudinal modes, as indicated by the existence of microinertia $j$ in both terms of the eigenfrequencies. This is in stark contrast to the dispersion relations of conventional elastic waves that are dominated by translation motion. In addition, we see that the chiral parameters $B_2$ and $B_3$ induce the splitting of the transverse modes.

By fitting the analytical dispersion relations and the constitutive relations with the numerical results of band structure and eigenmodes (see Supplementary Information), we retrieved the effective constitutive tensors $B_{ijkl}$, $C_{ijkl}$ and $D_{ijkl}$. We then apply these tensors to analytically evaluate the dispersion relations, and the results are plotted as the solid red lines in Fig. 3(c). In addition, we numerically simulated the band structures of the micropolar effective medium, and results are shown as the green markers in Fig. 3(c). All results agree excellently for $k < 0.15\pi/a$, demonstrating the validity of the effective medium description based on micropolar elasticity.

Under the effective medium description, the transverse modes are circularly polarized plane waves propagating in a homogeneous micropolar medium with acoustic activity. They carry well-defined spin and allows the possibility of achieving SOIs. In what follows, we demonstrate two SOI phenomena via numerical simulations and experiments. Effective medium theory based on micropolar elasticity is also applied to understand the results.

**Spin-orbit interaction in momentum space**

The transverse sound near the Γ point in Fig. 3(c) can be described by an effective Hamiltonian $H = -v\mathbf{S}\cdot\mathbf{k}$ with $\mathbf{S}$ being the spin-1 operator defined as $(S_i)_{jk} = -i\epsilon_{jki}$. The Hamiltonian indicates a coupling between the spin and linear crystal momentum $\mathbf{k}$, which induces splitting of the eigenfrequencies $\Delta\omega \propto k$ and leads to a "negative band" for the LCP sound with spin $s = \langle \text{LCP}|\mathbf{S}\cdot\hat{\mathbf{k}}|\text{LCP}\rangle = +1$, as shown in Fig. 3(c). Near the Γ point, the group velocity and phase velocity take opposite signs, indicating negative refraction for a sound wave passing the metamaterial-air interface. We can define an effective refractive index $n = -v_0 k/\omega_T^-$ with $v_0$ being the speed of sound in air [27]. This acoustic-activity-induced negative index is different from those derived from overlapped monopolar and dipolar resonances[42,43] or from multipole scattering[44]. It was proposed and verified in optics[27,28], but has been long considered impossible for sound, since longitudinal waves cannot distinguish material chirality. Next, we numerically and experimentally demonstrate negative refraction in the 3D micropolar metamaterial.

In the numerical simulation, we consider the metamaterial consisting of 5 unit cells along $z$ direction and 30 unit cells along $x$ direction, as shown in Fig. 4(a). A periodic boundary condition is applied in $y$ direction. A Gaussian beam obliquely incidents on the metamaterial at 70 degrees. Because the sound beam is longitudinal in the air but transverse in the metamaterial, impedance mismatch happens at the interfaces. For the efficient excitation of transverse sound, we engineered the surface impedance by adding acoustic tubes (see Supplementary Information). This also guarantees that only the $s = +1$ sound is excited in the metamaterial. Figures 4(a) and 4(b) show the real part and the amplitude of the pressure field, respectively. The negative refraction is clearly observed. To verify the effective-medium description of this phenomenon, we apply the effective parameters (same as those in Fig. 3(c)) to simulate the propagation of the same Gaussian beam in the micropolar effective medium. Negative refraction is seen again, as shown in Fig. 4(c).

Experimentally, we fabricated a smaller sample consisting of 11 × 4 unit cells, as shown in Fig. 5(a). This one-layer metamaterial can also induce negative refraction, as expected from the band structure of the 1D lattice system in Fig. 2(d). We indeed observed the phenomenon by measuring the transmitted pressure field in the yellow zone of Fig. 5(a). Figures 5(b) and (c) respectively show the amplitude and the real part of the pressure field. The beam with an incident angle of 40 degrees is generated by an array of speakers. The simulation results are shown in Fig. 5(d) and (e), where the region of experimental measurement is marked by the rectangle. Good agreement between the simulation and experimental results is seen, which confirms the negative refraction phenomenon induced by SOI.

**Spin-orbit interaction in real space**

The SOIs of transverse waves can also happen in real space. One intriguing phenomenon induced by such SOIs is the spin-dependent vortex generation in the scattering of subwavelength particles, which leads to the conversion of spin to OAM with important applications in optics such as optical manipulations and imaging[2,45–47]. It is commonly believed that airborne sound does not have this remarkable property. Here, we demonstrate the real-space SOI for the transverse sound in the micropolar metamaterial.

We consider the micropolar metamaterial consisting of 19 × 19 × 4 unit cells under the normal incidence of a Gaussian beam at $f = 655$ Hz (corresponding to the frequency of the "negative band"), as shown in Fig. 6(a). We remove one unit cell from the center of the metamaterial to create a subwavelength defect, as shown by the blue cube in Fig. 6(b). This defect then serves as an acoustic dipole particle. Figure 6(c) shows the amplitude of the transmitted pressure field obtained by simulations. We notice a spiral pattern with two arms, which is a signature of an optical vortex with topological charge $q = +2$. This phenomenon can be understood as a result of SOI mediated by the dipole particle. The longitudinal sound in air excites the transverse sound in the metamaterial that carries spin $s = +1$. The transverse sound has a velocity field $\mathbf{v}_0$ and a negative wave vector $\mathbf{k}_0$. It is scattered by the dipole particle, which generates scattered fields $\mathbf{v}_s$ with a negative wave vector $\mathbf{k}$, as shown in Fig. 6(b). The scattered field can be considered a spherical projection of the incident field: $\mathbf{v}_s \propto -\hat{\mathbf{r}} \times (\hat{\mathbf{r}} \times \mathbf{v}_0)$, where $\hat{\mathbf{r}}$ is the unit radial vector. The projection induces noncommutative SO(3) rotations of the incident field and leads to geometric phases that account for the spin-to-OAM conversion[5]. This process can be expressed as $|s\rangle \rightarrow c_1|s\rangle + c_2 e^{2is\varphi}|-s\rangle$, where $\varphi$ is the azimuthal angle, $c_1$ and $c_2$ are the coefficients characterizing the efficiency of the SOI[48]. The second term indicates the flip of spin and the emergence of an optical vortex with topological charge $q = 2s$. At the output interface, the background Gaussian beam and the scattered field are both converted to longitudinal sound, and their interference gives rise to the spiral pattern of pressure amplitude shown in Fig. 6(c). To verify the results, we simulate the phenomenon in the micropolar effective medium using the same effective parameters as in Fig. 4. Similar interference pattern of the velocity field is obtained inside the micropolar effective medium, as shown in Fig. 6(d). Figure 6(e) shows the real part of the scattered velocity field with $s = -1$, which clearly shows a $4\pi$ phase variation in the azimuthal direction and confirms the optical vortex with charge $q = +2$.

**Discussion**

We have demonstrated a mechanism that transforms airborne sound into a transverse wave with rich phenomena of SOIs. The SOIs are in contrast to the pseudo-SOIs in acoustic topological insulators where hybridization of modes are employed to construct "pseudo-spins"[49]. Our idea relies on engineering acoustic resonances at the subwavelength level to emulate shear responses, thereby giving rise to a fully vectorial transverse sound that carries a spin. From a microscopic perspective, this mechanism is similar to the emergence of induced dipole moments in a dielectric medium. Notably, dipole responses have been widely leveraged for anomalous effective mass density[50]. However, those dipole moments are parallel to the propagation direction, whereas in our micropolar metamaterial, the dipoles undergo microrotation in the plane orthogonal to the propagation direction. Consequently, a total of six degrees of freedom are needed to fully characterize the transverse sound in 3D, thereby bringing richer functionalities for sound manipulations. In particular, the emergence of acoustic activity can enable chiral sound-matter interactions with unprecedented applications, such as chiral discrimination and sensing, acoustic manipulations of chiral particles, and acoustic circular dichroism, etc. The spin-1 sound demonstrated here can also realize the bosonic analogue of Kramers doublet. We thus expect a variety of applications and extensions of the results in spin-orbit acoustics, topological acoustics, and novel acoustic metamaterials.

## Methods

**Micropolar effective medium theory**

Near the $\Gamma$ ($k = 0$) point, the acoustic metamaterial is approximately equivalent to a homogeneous and isotropic micropolar medium. Each point of the medium is characterized by a displacement field **u** and a microrotation field **ϕ**. Using Einstein summation convention, the strain tensor and curvature tensor can be expressed as[22]

$$\varepsilon_{ij} = \frac{\partial u_j}{\partial x_i} - \epsilon_{ijk}\phi_k, \tag{1}$$

$$\kappa_{ij} = \frac{\partial \phi_j}{\partial x_i}. \tag{2}$$

The constitutive relations are [25,34]

$$\sigma_{ij} = C_{ijkl}\varepsilon_{kl} + B_{ijkl}\kappa_{kl}, \tag{3}$$

$$m_{ij} = B_{klij}\varepsilon_{kl} + D_{ijkl}\kappa_{kl}, \tag{4}$$

where the elastic constitutive tensors can be expressed as

$$\begin{aligned}
C_{ijkl} &= C_1\delta_{ij}\delta_{kl} + C_2\delta_{ik}\delta_{jl} + C_3\delta_{il}\delta_{jk}, \\
B_{ijkl} &= B_1\delta_{ij}\delta_{kl} + B_2\delta_{ik}\delta_{jl} + B_3\delta_{il}\delta_{jk}, \\
D_{ijkl} &= D_1\delta_{ij}\delta_{kl} + D_2\delta_{ik}\delta_{jl} + D_3\delta_{il}\delta_{jk}.
\end{aligned} \tag{5}$$

In terms of conventional notation, we have

$$\begin{aligned}
C_1 &= \lambda, C_2 = \mu + \kappa, C_3 = \mu - \kappa, \\
B_1 &= \eta, B_2 = \zeta + \xi, B_3 = \zeta - \xi, \\
D_1 &= \alpha, D_2 = \beta + \gamma, D_3 = \beta - \gamma.
\end{aligned} \tag{6}$$

Here, $\lambda$ and $\mu$ are the Lame constants; $\kappa$, $\alpha$, $\beta$ and $\gamma$ are the micropolar elastic constants; $\eta, \zeta$ and $\xi$ are the elastic constants due to material chirality. The equations governing the propagation of the sound wave in the chiral micropolar medium are given by the conservation of linear momentum and angular momentum:

$$\frac{\partial \sigma_{ji}}{\partial x_j} = \rho \frac{\partial^2 u_i}{\partial t^2}, \tag{7}$$

$$\frac{\partial m_{ji}}{\partial x_j} + \epsilon_{ijk}\sigma_{jk} = j\frac{\partial^2 \phi_i}{\partial t^2}, \tag{8}$$

where $\rho$ is the mass density and $j$ is the microinertia density. Assume time-harmonic forms of the displacement field $u_i = U_i e^{ik_i x_i - i\omega t}$ and microrotation field $\phi_i = \Phi_i e^{ik_i x_i - i\omega t}$ and use the constitutive relations, the above governing equations can be reduced to

$$-k_j k_k C_{jikl} U_l + (ik_j \epsilon_{nkl} C_{jikn} - k_j k_k B_{jikl})\Phi_l = -\rho \omega^2 U_i, \tag{9}$$

$$\begin{aligned}(ik_n \epsilon_{ijk} C_{jknl} - k_j k_k B_{klji})U_l + ik_j(\epsilon_{nkl} B_{knji} + \epsilon_{ink} B_{nkjl})\Phi_l \\ -(k_j k_k D_{jikl} - \epsilon_{ijk}\epsilon_{nml} C_{jkmn})\Phi_l = -j\omega^2 \Phi_i.\end{aligned} \tag{10}$$

Express $U_i$ in terms of $\Phi_i$ by using Eq. (9) and substitute it into Eq. (10), we obtain

$$H\boldsymbol{\Phi} = [-v\mathbf{S}\cdot\mathbf{k} + a_1 \mathbf{k}\mathbf{k} + a_2 k^2 + O(k^3)]\boldsymbol{\Phi} = \delta\omega\boldsymbol{\Phi}, \tag{11}$$

where we have expanded the equation at $k \to 0$ and $\omega - \omega_0 = \delta\omega \to 0$ with $\omega_0 = \sqrt{2(C_2 - C_3)/j}$. Here, $H$ is the effective Hamiltonian, $\mathbf{S}$ is the spin-1 matrix operator defined as $(S_i)_{jk} = -i\epsilon_{jki}$, $v = (B_2 - B_3)/\sqrt{2j(C_2 - C_3)}$, $a_1 = (D_1 + D_3)/\sqrt{8j(C_2 - C_3)} - \sqrt{2j(C_2 - C_3)}/8\rho$ and $a_2 = D_2/\sqrt{8j(C_2 - C_3)} + \sqrt{2j(C_2 - C_3)}/8\rho$. It is noted that the leading order of the effective Hamiltonian describes the spin-orbit interaction. The above equation gives three eigenmodes that are dominated by the microrotation of mass points, among which two are transverse waves and one is a longitudinal wave. Their dispersion relations (retained the lowest order of $k$) are

$$\omega_T^\pm = \omega_0 \pm vk \,, \omega_L = \omega_0 + \tau k^2, \tag{12}$$

where $\tau = (D_1 + D_2 + D_3)/\sqrt{8j(C_2 - C_3)}$. In the low-frequency limit, microrotation vanishes in the metamaterial due to the cut-off frequencies of the resonators. Thus, $\mu, \kappa, \alpha, \beta, \gamma, \eta, \zeta$ and $\xi$ all vanish, only $C_1 = \lambda$ (i.e. bulk modulus) remains. In this case, the metamaterial reduces to conventional acoustic metamaterial without bianisotropy.

**Effective parameters retrieval**

We retrieved the effective parameters based on the numerically computed band structures and the eigenmodes. Among the total 11 material parameters, only 9 parameters (i.e. $B_2, B_3, C_2, C_3, D_1, D_2, D_3, \rho, j$) contribute to the microrotation-dominated waves that are responsible for the SOI phenomena. $B_1$ and $C_1$ do not play a role in the effective properties of the metamaterial for these waves. A three-step approach is applied to retrieve the effective parameters. We first evaluated the total force and torque acting on the unit cell and applied Newton's 2nd law to calculate the effective mass density $\rho$ and microinertia density $j$. Then, we fit the analytical dispersion relations with high-order corrections to the numerically computed band structures, from which the values of $C_2 - C_3, B_2, B_3, D_2, D_1 + D_3$ can be determined. To further determine the values of $C_2, C_3, D_1$ and $D_3$, we employ the constitutive relations of Eqs. (3) and (4), where the strain and coupling stress can be obtained via boundary averaging of the eigenmode fields. The details about the parameter retrieval can be found in Supplementary Information. For a narrow frequency region near the Γ point, the retrieved effective parameters are approximately constants: $\rho = 0.637 \text{ kg/m}^3$, $j = 5.64 \times 10^{-4} \text{ kg/m}$, $B_2 = 5.91 \text{ N/m}$, $B_3 = 55.0 \text{ N/m}$, $C_2 =$ $-1.68 \times 10^4$ Pa, $C_3 = -2.16 \times 10^4$ Pa, $D_1 = 20.3$ N, $D_2 = 2.69$ N, $D_3 = -16.2$ N. These material parameters are then used in full-wave numerical simulations of the micropolar effective medium to verify the band structures and SOI phenomena of the metamaterial systems.

**Numerical simulations**

Full-wave numerical simulations are performed by using the finite-element package COMSOL Multiphysics[51] for both the metamaterial systems and the micropolar effective medium. For the resonators in Figs. 1, 3, 4, and 6, we set the radius $R = 5$ cm and height $h = 2$ cm. The period in both 1D and 3D metamaterials is $a = 12.1$ cm. The tubes have radii $r = 0.2$ cm. For the chiral resonator, the upper surface is twisted $\pi/2$ with respect to the bottom surface. The Gaussian beam in Figs. 4 and 6 has a beam width $w = 1.2\lambda$. A sound-hard boundary condition is applied on all the boundaries of the resonators and tubes. Floquet periodic boundary conditions are applied to the 1D lattice and 3D metamaterials to compute the band structures. To compute the band structures of the micropolar effective medium and to simulate the associated SOIs phenomena, we developed weak-form formulations for the micropolar constitutive relations and momentum conservation equations, which are then implemented using COMSOL. The band structures of the micropolar medium are calculated by considering a unit cell made of homogenous and isotropic micropolar medium with the retrieved effective parameters.

**Experiments**

The 1D lattice and 3D metamaterial were fabricated by using 3D printing. The resonators and connecting tubes are made of acrylonitrile butadiene styrene (ABS plastics), which were then assembled to form the structures in Fig. 2(a) and Fig. 5(a). The fabricated units correspond to a scaled version of the units in Fig. 1(a)(c) and Fig. 3(a) with $R = 3.5$ cm, $h = 1.75$ cm, $r = 0.4$ cm and $a = 10.7$ cm. For the band structures of the 1D lattices, we excite the lattice using a loudspeaker at one end. The signal is generated by a waveform generator (Keysight 33500B) as a short pulse covering the frequency range of interest. We then measure the pressure responses with a microphone and a digital oscilloscope (Keysight DSO2024A) at all 24 unit cells with one measurement point per cell. Then, we perform a 2D Fourier transform to obtain the dispersion curves, which show the band structures (Fig. 2(c, d)). For the negative refraction experiment, we used an array of 11 loudspeakers to generate an obliquely incident Gaussian beam. Each speaker was driven by an independent channel of a computer sound interface (MOTU 16A). Both the amplitudes and phases of the output signal from each channel were precisely controlled by a PC (via a MATLAB program) to generate the targeted Gaussian beam with a tilted phase profile in order to emulate oblique incidence at the chosen angle. On the far side of the metamaterial, small horns are connected to each unit cell to improve impedance matching between the metamaterial and the air. The tabletop and a top plate (removed in Fig. 5(a) to show the metamaterials) form a 2D waveguide in the output region for the better observation of the negative refracted field profile. A microphone is carried by a translational stage to raster-map the output beam profiles.

## Acknowledgements

The work described in this paper was supported by grants from the Research Grants Council of the Hong Kong Special Administrative Region, China (Projects No. CityU 21302018 and No. C6013-18G). G. M. is supported by National Natural Science Foundation of China Excellent Young Scientist Scheme (Hong Kong and Macau) (No. 11922416) and Youth Program (No. 11802256). We thank Profs. C. T. Chan, Z. Q. Zhang, Y. Wu and Dr. R. Y. Zhang for helpful discussions and Dr. Y. Chen for assistance in the implementation of weak form.

## Author contributions

S.W. conceived the idea and designed the metamaterial and conducted the numerical simulations. G.Z. and Q.T. assisted in numerical simulations. S.W. and Q.T. developed the effective medium theory. G.Z. and X.W. performed the experiments. J.L. assisted in theoretical interpretations. S.W. wrote the manuscript with input from all authors. S.W. and G.M. supervised the project. All authors contributed to the data analysis and the polishing of the manuscript.

## Competing interests

The authors declare no competing interests.

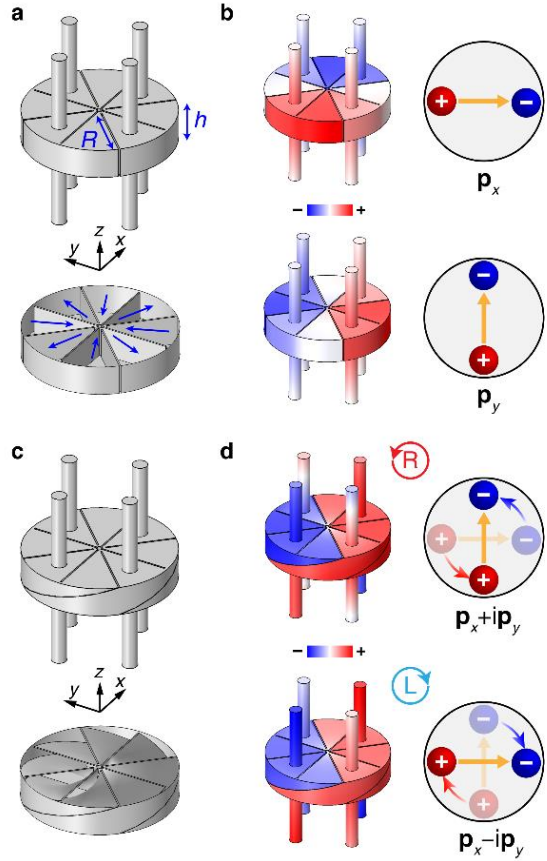

**Figure 1. Eigenmodes of the 1D acoustic lattices.** (a) The unit cell of the achiral lattice. The arrows show the flow of air inside the resonator. (b) The pressure eigenfields of the two transverse dipole modes. The velocity is linearly polarized on the transverse plane, corresponding to acoustic dipoles $\mathbf{p}_x$ and $\mathbf{p}_y$. The positive (negative) charge corresponds to positive (negative) pressure. The yellow arrows denote the velocity field. (c) The unit cell of the chiral lattice. (d) Pressure eigenfields of the chiral dipole modes. The velocity fields are circularly polarized on the transverse plane, corresponding to circularly polarized dipoles $\mathbf{p}_x \pm i\mathbf{p}_y$.

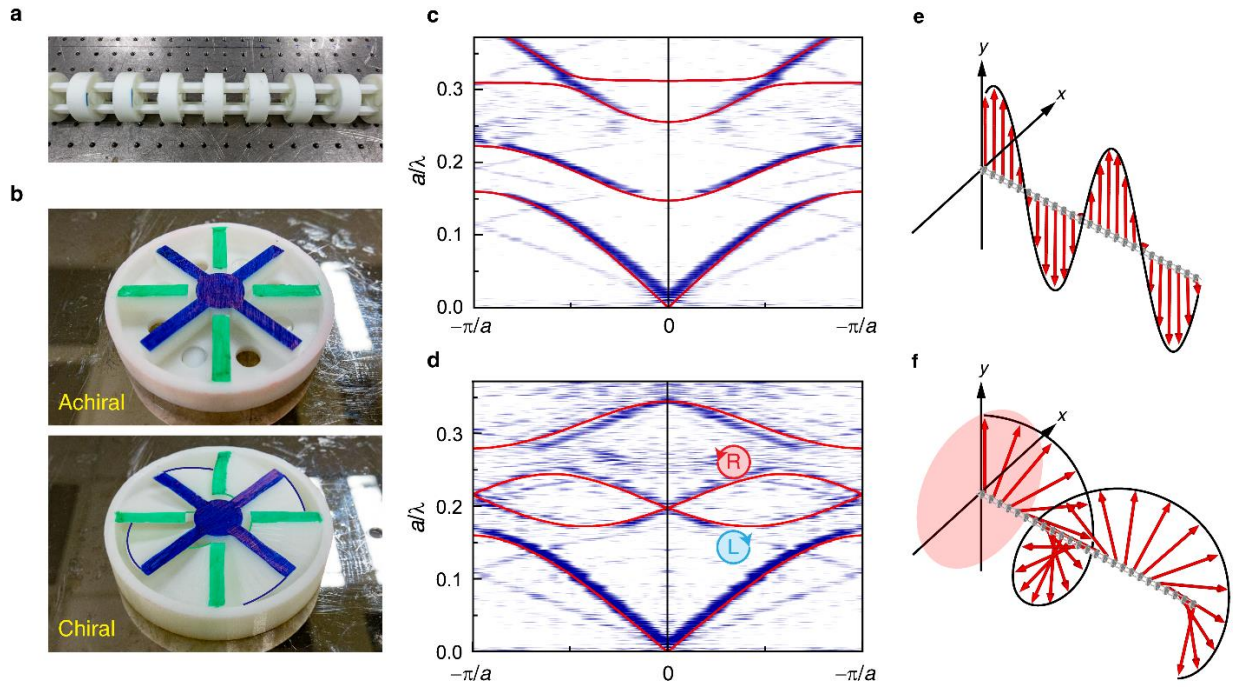

**Figure 2. Band properties of the 1D acoustic lattices.** (a) The fabricated sample of the 1D lattice. (b) The internal structures of the achiral and chiral unit cells. Experimental (blue) and numerical (red lines) results for the band structure of the achiral lattice (c) and the chiral lattice (d). The averaged velocity field of the achiral lattice shows a linear polarization (e), whereas circular polarization is seen for the chiral lattice (f).

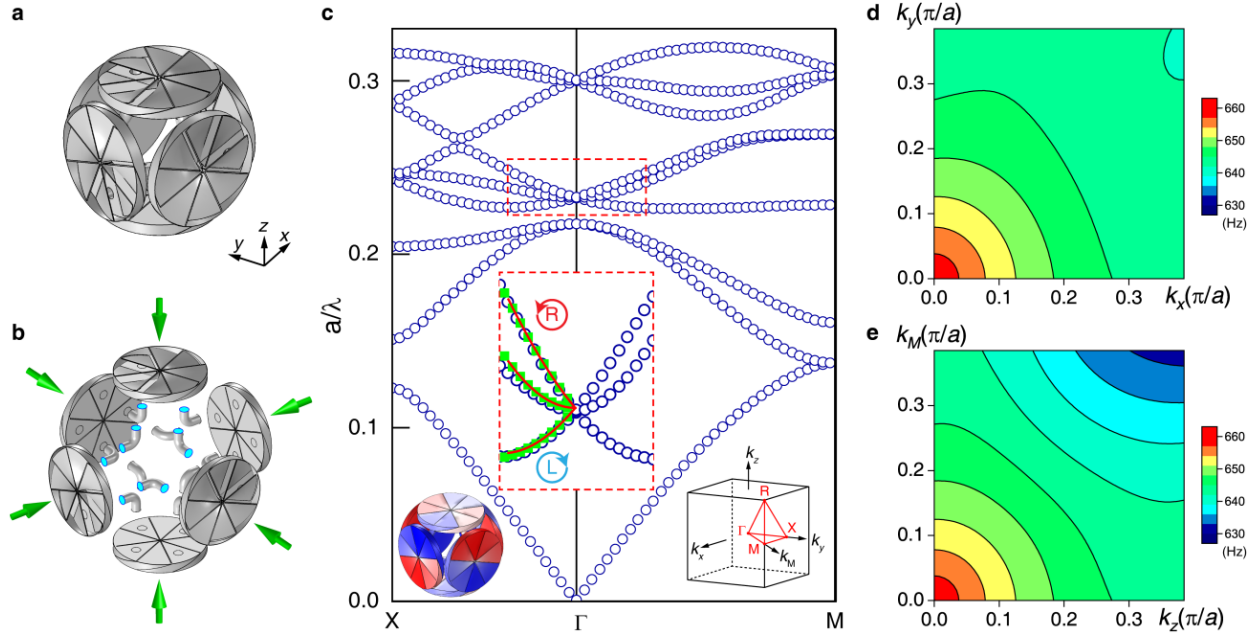

**Figure 3. The 3D acoustic micropolar metamaterials.** (a) The unit cell consists of 3 orthogonally arranged resonators connected with tubes. (b) The details of the unit-cell components. (c) The band structure of the metamaterial. The middle inset (dashed red box) shows a zoom-in of the dipole bands. The blue circles denote the numerical results of the metamaterial. The green squares denote the numerical results of the micropolar effective medium. The solid red lines denote the analytical results. The inset at the left corner shows the pressure eigenfield of the LCP mode. The inset at the right corner shows the Brillouin zone. (d) The isofrequency contours of the negative band in $k_x$-$k_y$ plane. (e) The isofrequency contours of the negative band in $k_z$-$k_M$ plane.

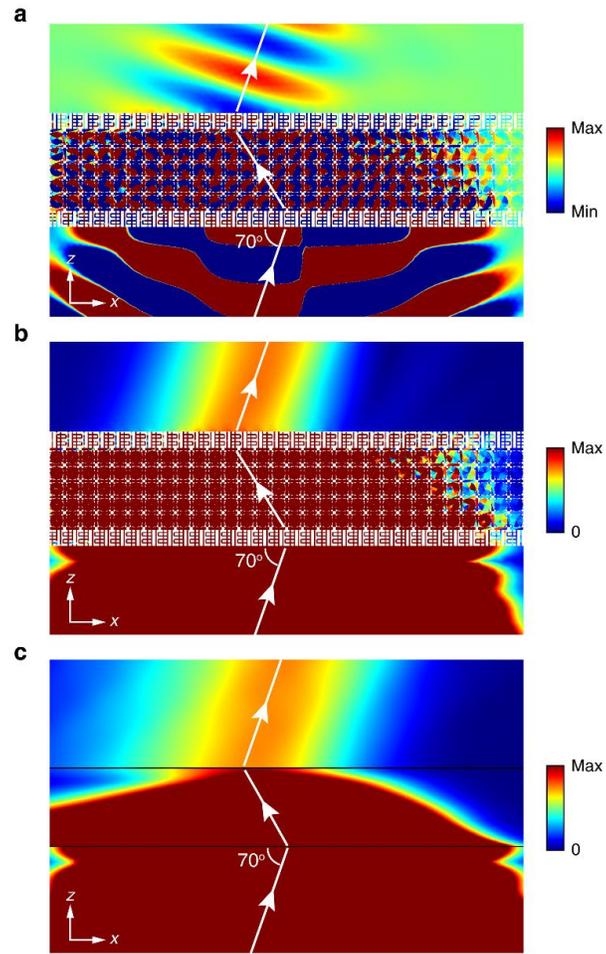

**Figure 4. Negative refraction induced by SOI in momentum space.** The real part (a) and the amplitude (b) of the pressure field for a Gaussian beam that propagates through the acoustic metamaterial with an incident angle of 70 degrees. (c) The amplitude of pressure field in the corresponding micropolar effective medium system.

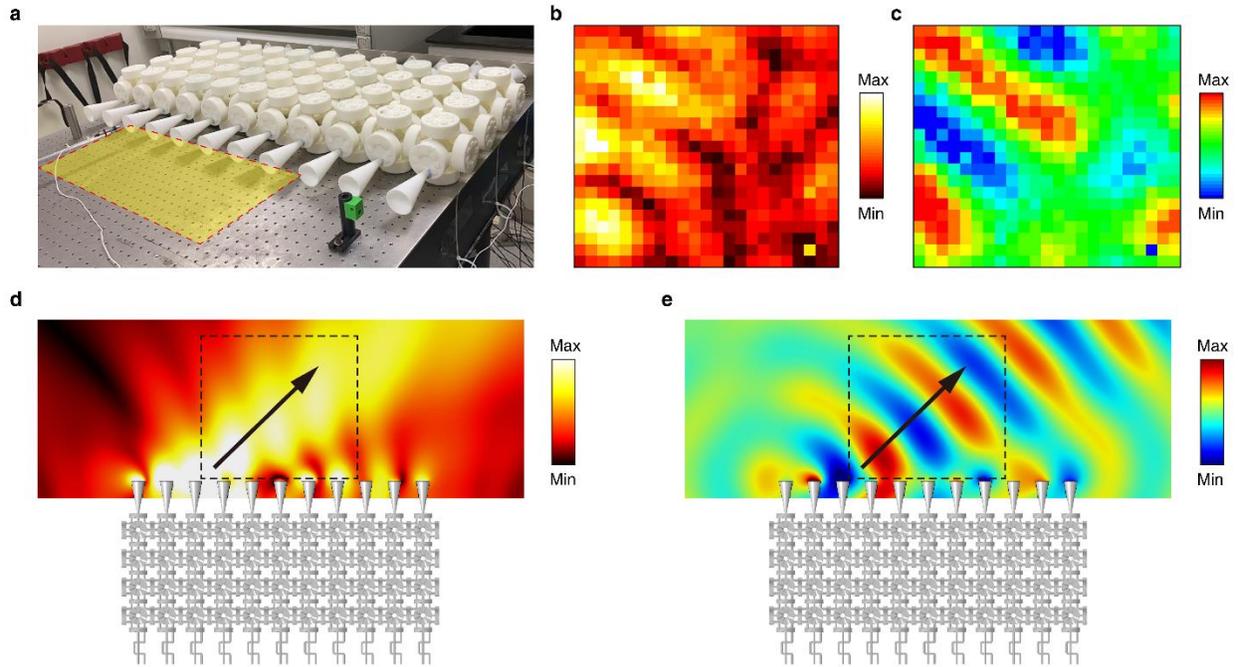

**Figure 5. Experimental demonstration of the negative refraction.** (a) A photograph of the metamaterial lattice and the measurement area (yellow colored). (b) The amplitude and (c) the real part of the measured pressure field. (d) The amplitude and (c) the real part of the simulated pressure field. The dashed boxes indicate the corresponding measurement area in the experiment.

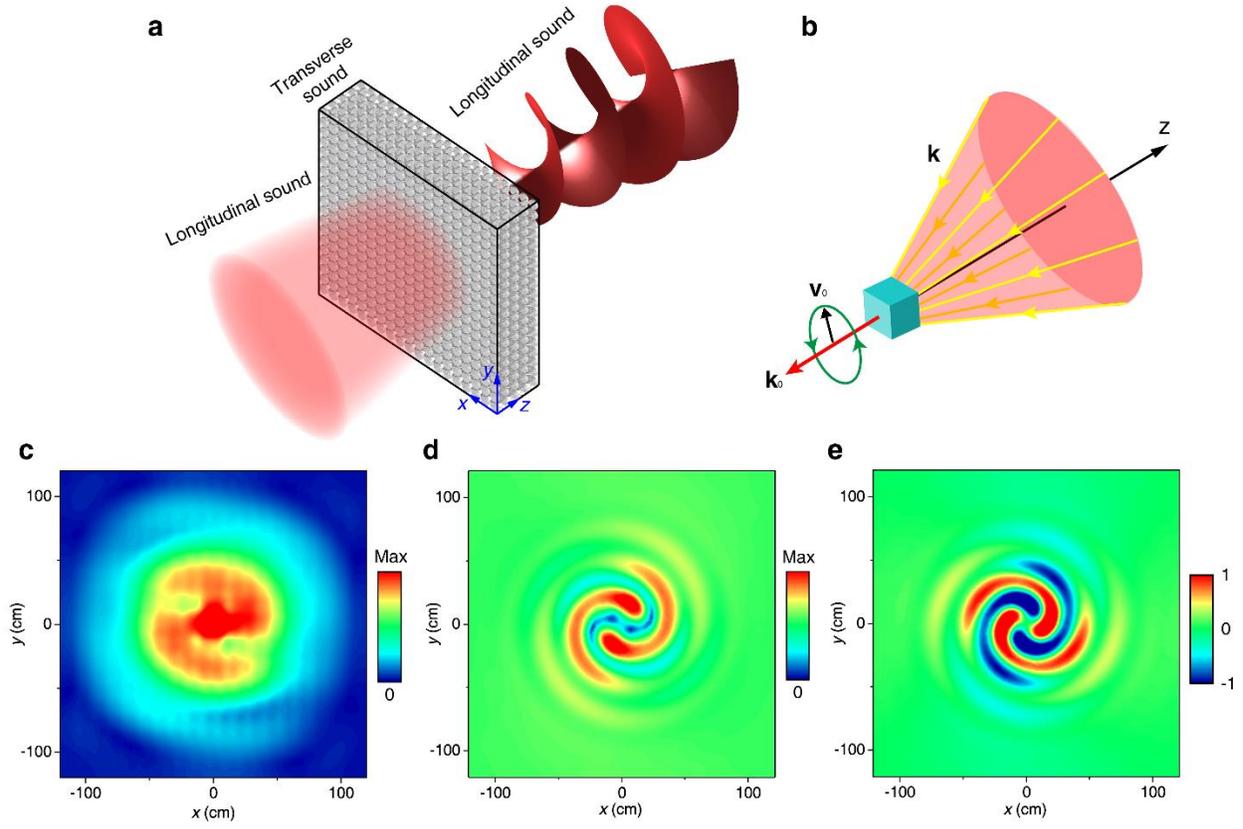

**Figure 6. Spin-dependent vortex generation enabled by SOI in real space.** (a) The schematic of the scattering system. One unit cell is removed from the center of the metamaterial to create a dipole scatterer. A Gaussian beam is normally incident on the metamaterial. (b) The schematic of the scattering of transverse sound inside the metamaterial. The blue cube denotes the scatterer. (c) The amplitude of the transmitted pressure field. (d) The velocity amplitude in the micropolar effective medium due to the interference of $s = -1$ scattered field with the background field. (e) The real part of the $s = -1$ scattered velocity field in the micropolar effective medium.

Supplementary Information for

# Spin-orbit interactions of transverse sound

Shubo Wang[1,*], Guanqing Zhang[2], Xulong Wang[2], Qing Tong[1], Jensen Li[3], and Guancong Ma[2,*]

[1] Department of Physics, City University of Hong Kong, Tat Chee Avenue, Kowloon, Hong Kong, China

[2] Department of Physics, Hong Kong Baptist University, Kowloon Tong, Hong Kong, China

[3] Department of Physics, The Hong Kong University of Science and Technology, Clear Water Bay, Hong Kong, China

*Correspondence should be addressed to:
Shubo Wang (shubwang@cityu.edu.hk)
Guancong Ma (phgcma@hkbu.edu.hk)

## Supplementary Note 1: Retrieval of effective material parameters

To characterize the material properties of the micropolar metamaterial, we need 9 parameters (i.e., $B_1, B_2, B_3, C_1, C_2, C_3, D_1, D_2, D_3$) to describe the elasticity and two parameters (i.e., mass density $\rho$ and microinertia density $j$) for the dynamic equations. Here, we apply a three-step method to determine the values of these parameters. The retrieved parameters are then used for numerical simulations of the micropolar effective medium. A comparison between the results of the effective medium and that of the metamaterial can validate our understanding of the metamaterial's acoustic properties. Since the metamaterial is approximately isotropic near Γ point (as seen from the isofrequency contours shown in Fig. 3(d) and 3(e) of the main text), we consider the LCP sound with $s = +1$ propagating in the $+x$ direction and use the eigenfields in the following evaluations to obtain the effective parameters.

The mass density $\rho$ can be determined via Newton's 2$^{nd}$ law

$$\rho = \frac{m}{V} = \frac{F_x}{\ddot{u}_x a^3} = \frac{-F_x}{a^3 \omega^2 u_x}. \tag{S1}$$

Here $a$ is the size of the unit cell, $F_x$ is the $x$ component of the net force acting on the unit cell, which can be calculated as $F_x = \hat{\mathbf{x}} \cdot \int \hat{\mathbf{n}} p dS$, $\hat{\mathbf{x}}$ is the unit vector of $x$ axis, $\hat{\mathbf{n}}$ is the inward normal unit vector, $p$ is the pressure, and the integral is evaluated over all surfaces of the unit cell. $u_x$ is the averaged displacement of the unit cell along $x$ direction that is defined as $u_x = \frac{1}{6a^2}\hat{\mathbf{x}} \cdot \int \mathbf{u} dS$, and the integral is evaluated over all 6 surfaces of the unit cell (corresponding to the blue-colored surfaces in Fig. S1).

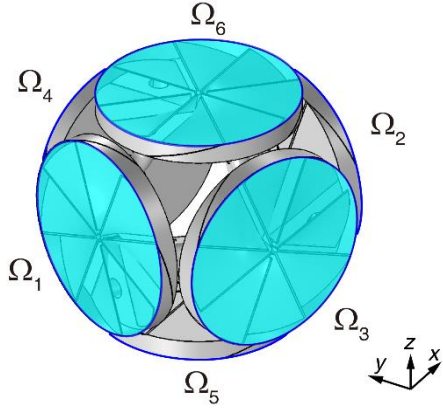

Fig. S1. Unit cell of the micropolar metamaterial.

The micro-rotation inertia density $j$ can be determined as

$$j = \frac{J}{V} = \frac{1}{a^3}\sum_{i=1}^{3}\frac{T_x^i}{\ddot{\Phi}_x^i} = \frac{-1}{a^3\omega^2}\sum_{i=1}^{3}\frac{T_x^i}{\Phi_x^i}, \tag{S2}$$

where $T_x^i = \hat{\mathbf{x}} \cdot \int (\mathbf{r} - \mathbf{r}_i) \times \hat{\mathbf{n}} p dS$ is the $x$-component of the torque exerted on $i$th resonator (each unit cell has 3 resonators) with respect to its center $\mathbf{r}_i$, $\Phi_x^i = \frac{1}{V}\hat{\mathbf{x}} \cdot \int \frac{(\mathbf{r}-\mathbf{r}_i)\times \mathbf{u}}{|\mathbf{r}-\mathbf{r}_i|^2} dV$ is the average rotation of the volumetric air inside the resonator with respect to the center, and $V$ is the volume of the resonator.

In the second step, we employ the dispersion relations to retrieve part of the constitutive tensor parameters. The dispersion relations Eq. (12) in the main text are derived by keeping the leading orders up to $O(k^2)$, which already can capture the underlying physics and reproduce the key features of the band structures. To use it for parameter retrieval, we adopt the following dispersion relations with leading orders up to $O(k^3)$.

$$\omega_T^\pm = \sqrt{\frac{2(C_2 - C_3)}{j}} \pm \frac{B_2 - B_3}{\sqrt{2j(C_2 - C_3)}}k + \frac{(C_2 - C_3)^2 j - [(B_2 - B_3)^2 - 2(C_2 - C_3)D_2]\rho}{4\rho\sqrt{2j(C_2 - C_3)^3}}k^2$$
$$\pm \frac{(B_2 + 3B_3)(C_2 - C_3)^2 j + (B_2 - B_3)[(B_2 - B_3)^2 - 2(C_2 - C_3)D_2]\rho}{8\rho\sqrt{2j(C_2 - C_3)^5}}k^3, \tag{S3}$$

$$\omega_L = \sqrt{\frac{2(C_2 - C_3)}{j}} + \frac{D_1 + D_2 + D_3}{2\sqrt{2j(C_2 - C_3)}}k^2, \tag{S4}$$

The above equations are fitted to the numerically obtained band structures in Fig. 3(c). The fitting can give the values of $B_2, B_3, D_2, C_2 - C_3$ and $D_1 + D_3$. We note that $B_1$ and $C_1$ do not appear in the above dispersion relations, which indicates that their contribution to the considered transverse modes is negligible. This is confirmed by numerical simulations of the micropolar effective medium.

To determine the values of $C_2, C_3, D_1$ and $D_3$, we then employ the constitutive relations in Eqs. (3) and (4) of the main text. The following two equations are sufficient to determine their values:

$$m_{12} = B_2 \varepsilon_{12} + B_3 \varepsilon_{21} + D_2 \kappa_{12} + D_3 \kappa_{21}, \tag{S5}$$

$$\sigma_{12} = C_2 \varepsilon_{12} + C_3 \varepsilon_{21} + B_2 \kappa_{12} + B_3 \kappa_{21}, \tag{S6}$$

where the couple stress $m_{12}$ is evaluated as

$$m_{12} = \frac{1}{2a^2} \hat{\mathbf{y}} \cdot \left[ \int_{\Omega_2} (\mathbf{r} - \mathbf{r}_i) \times \hat{\mathbf{n}} p \, dS - \int_{\Omega_1} (\mathbf{r} - \mathbf{r}_i) \times \hat{\mathbf{n}} p \, dS \right], \tag{S7}$$

where "$\Omega_1$" and "$\Omega_2$" denote the surfaces of the two half resonators as labeled in Fig. S1. The force stress $\sigma_{12}$ is evaluated as

$$\sigma_{12} = \frac{1}{2L} \hat{\mathbf{y}} \cdot \left[ \int_{x=+\frac{a}{2}} \hat{\mathbf{n}} p \, dl - \int_{x=-\frac{a}{2}} \hat{\mathbf{n}} p \, dl \right], \tag{S8}$$

where the two integrals are evaluated over all edges of the boundary surface at $x = +a/2$ and $x = -a/2$, respectively. $L$ is the total length of the edges of each boundary surface. The strain $\varepsilon_{12}$ and $\varepsilon_{21}$ are evaluated as

$$\varepsilon_{12} = \frac{1}{a^3} \hat{\mathbf{y}} \cdot \left[ \int_{x=+\frac{a}{2}} \mathbf{u} \, dS - \int_{x=+\frac{a}{2}} \mathbf{u} \, dS \right], \varepsilon_{21} = \frac{1}{a^3} \hat{\mathbf{x}} \cdot \left[ \int_{y=+\frac{a}{2}} \mathbf{u} \, dS - \int_{y=+\frac{a}{2}} \mathbf{u} \, dS \right] \tag{S9}$$

and the curvature $\kappa_{12}$ and $\kappa_{21}$ are evaluated as

$$\kappa_{12} = \frac{2}{aV} \hat{\mathbf{y}} \cdot \left[ \int_{\Omega_2} \frac{(\mathbf{r} - \mathbf{r}_i) \times \mathbf{u}}{|\mathbf{r} - \mathbf{r}_i|^2} dV - \int_{\Omega_1} \frac{(\mathbf{r} - \mathbf{r}_i) \times \mathbf{u}}{|\mathbf{r} - \mathbf{r}_i|^2} dV \right], \tag{S10}$$

$$\kappa_{21} = \frac{2}{aV}\hat{\mathbf{x}} \cdot \left[ \int_{\Omega_4} \frac{(\mathbf{r} - \mathbf{r}_i) \times \mathbf{u}}{|\mathbf{r} - \mathbf{r}_i|^2} dV - \int_{\Omega_3} \frac{(\mathbf{r} - \mathbf{r}_i) \times \mathbf{u}}{|\mathbf{r} - \mathbf{r}_i|^2} dV \right]. \tag{S11}$$

where "$\Omega_3$" and "$\Omega_4$" denote the surfaces of the two half resonators as labelled in Fig. S1. Using Eqs. (S5)-(S11), we determined the values of the remaining effective parameters $C_2, C_3, D_1$ and $D_3$.

## Supplementary Note 2: Surface impedance engineering

Since the sound inside the micropolar metamaterial is a transverse wave while the sound in air is a longitudinal wave, impedance mismatch occurs at the air-metamaterial interface. Thus, it is difficult to excite the transverse wave by direct illumination of sound waves from air. The excitation efficiency can be improved by engineering the surface impedance of the metamaterial. As shown in Fig. S2, we tuned the length of the four connection tubes of the interface resonators to introduce a phase gradient of $(0, \pi/2, \pi, 3\pi/2)$ to the longitudinal input wave. This will preferably excite the LCP transverse sound that gives rise to the negative refraction and vortex generation. A similar design is applied to the output interface to improve the transmission from the metamaterial to the air. In the experiments, as shown in Fig. 5(a) and (d), we adopt a simplified design and only used two tubes of different length to couple the incident longitudinal sound into the metamaterial, and one horn-shaped tube is used at the output interface to improve the transmission of the transverse sound to air.

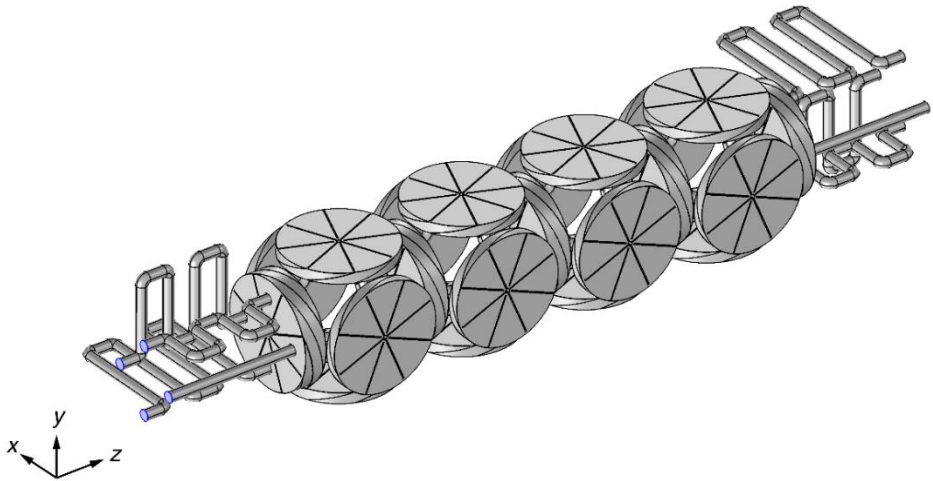

Fig. S2. Surface impedance engineering of the metamaterial.

4